\documentclass[prb,twocolumn,aps,showpacs]{revtex4}
\usepackage{graphicx,color}
\usepackage{latexsym,amssymb}

\begin{document}

\title{Narrow depression in the density of states at the Dirac point in disordered graphene}
\author{L. Schweitzer}
\affiliation{%
Physikalisch-Technische Bundesanstalt (PTB), Bundesallee 100, 38116 Braunschweig, Germany}

\begin{abstract}
The electronic properties of non-interacting particles moving on a
two-dimensional bricklayer lattice are investigated numerically. In
particular, the influence of disorder in form of a spatially varying random
magnetic flux is studied. In addition, a strong perpendicular constant
magnetic field $B$ is considered. The density of states $\rho(E)$ goes to zero 
for $E\to 0$ as in the ordered system, but with a much steeper slope. This
happens for both cases: at the Dirac point for $B=0$ and at the center
of the central Landau band for finite $B$. Close to the Dirac point,  
the dependence of $\rho(E)$ on the system size, on the disorder strength, and
on the constant magnetic flux density is analyzed and fitted to an analytical
expression proposed previously in connection with the thermal quantum Hall
effect. Additional short-range on-site disorder completely replenishes the
indentation in the density of states at the Dirac point.
\end{abstract}


\maketitle

\section{Introduction}
Despite enormous efforts in recent years, the electronic properties of
graphene near the neutrality point that separates conduction and valence bands 
are still under intense investigations. Theories aimed at describing graphene
via a Dirac-like equation have elucidated many intriguing effects like the
physics of massless relativistic particles and Klein
tunneling\cite{ANS98,KNG06,And07,Bee08,BTB09}, which has been experimentally
observed recently.\cite{YK09,SHG09} The Dirac-fermion approach, which
represents an approximation to the true lattice situation, is believed to be 
valid especially close to energy zero, the so called Dirac points, where the
conduction and valence bands touch each other and the dispersion is linear. 
Furthermore, nearest neighbor tight-binding descriptions,\cite{HHKM90,WFAS99,HFA06}
which particularly emphasize the hexagonal lattice structure of the carbon
sheet, have proven to be extremely helpful in understanding the basic
transport properties of this promising new material. 

Based on the observation of a peculiar quantum Hall
effect\cite{Nea05,ZTSK05,GZKPGM07}, it is generally accepted by now that some
kind of disorder must be present in the experimental setup. The disorder
influences the charge transport through the graphene sheet and affects the
measurable quantities at least quantitatively.\cite{Tea07} Yet, which type of
disorder is encountered in real samples is still completely unclear 
or only partly known in some special cases. This lack of knowledge is
particularly unfortunate as the definite type of disorder entirely determines 
the physical properties,\cite{SSW06,OGM06,KA07,NM07,OGM08,SM08a,KHA09} e.g.,
leading to complete Anderson localization in the case of short range
electrostatic scattering potentials via chiral symmetry breaking and
scattering between valleys.\cite{AE06,Alt06}       
For disordered systems, even the single particle density of states
(DOS) near the Dirac point remains still under debate. Depending on the
disorder type and approach, a vanishing, a finite, or an infinite DOS at the
Dirac point has been suggested for graphene or related 
models.\cite{LFSG94,NTW94,AHMZ00,RH01,PGC06,Khv08,Wea08,PLC08,HHS08,DZT08,ZDT09}  

Also, the interpretation of experimental results is hampered by the
uncertainty regarding the precise form of the DOS. Recently, following
earlier experimental investigations of the 
Landau level splitting in high magnetic fields,\cite{Zea06,JZSK07} the opening
of a spin (Zeeman) gap in the density of states at the Dirac point has been
suggested in the interpretation of magneto-transport measurements on graphene
sheets.\cite{GPNGKMZ09}  Only if a gap, separating electron and hole
states at the Dirac point, was assumed, the experimental data could be
accounted for. Another unexplained experimental observation to be found near the
Dirac point in the presence of a strong magnetic field is the divergent
resistance,\cite{CLO08,CLO09} which has been attracting considerable attention
lately. 

Theoretically, the opening of a mobility gap within the central Landau band
has been recently discovered by means of detailed two-terminal conductance
calculations.\cite{SM08a} It was found that with increasing disorder, the
critical energies where the plateau transitions of the Hall conductivity take
place, move apart. This splitting of the central conductance peak unveiled the
existence of an extra chiral quantum phase transition occurring at zero energy
with critical properties that differ from those of the quantum Hall
transitions.\cite{SM08a}  

In the present work, the single particle density of states is calculated
numerically for the same bricklayer lattice model. The presumed ripple
disorder is modeled by a spatially varying random magnetic flux with 
zero mean, pointing perpendicular to the two-dimensional lattice.
Also, an additional constant magnetic field is applied that leads
to the formation of Landau bands. It is shown that in the disordered case, the
density of states goes to zero at the Dirac point not only in the absence of a
perpendicular magnetic field.  Rather, a narrow suppression in the DOS is
obtained also in the presence of a finite magnetic field within the lowest
(central) Landau band. This unexpected feature depends essentially on the
disorder strength, on the system size, and on the strength of the
perpendicular constant magnetic field. Due to the neglect of electron spin in  
the model Hamiltonian, this outcome can not be attributed to a Zeeman
splitting, but must originate from chirality and a disorder-induced
interaction between the two sub-lattices. This consideration is confirmed by
the observation that the addition of short-range potential disorder completely
destroys the DOS-depression near the Dirac point.

\section{Bricklayer Model}
In the present study, the two-dimensional honeycomb lattice responsible for
the peculiar electronic properties of graphene is replaced by a bricklayer
model\cite{HHKM90,WFAS99,SM08a} which shares the same topology as the
hexagonal lattice. The bricklayer lattice is bi-partite and consists 
of two sub-lattices that can be constructed by rectangular unit cells of size
$2a\times a$ placed along the $x$-direction. The unit cell contains two sites
connected by a bond of length $a$. Each site on one sub-lattice is attached to 
three neighbors belonging to the other sub-lattice by two bonds in the
$\pm x$-direction and one alternating bond in the $\pm y$-direction. 

\begin{figure}
\includegraphics[width=7.25cm]{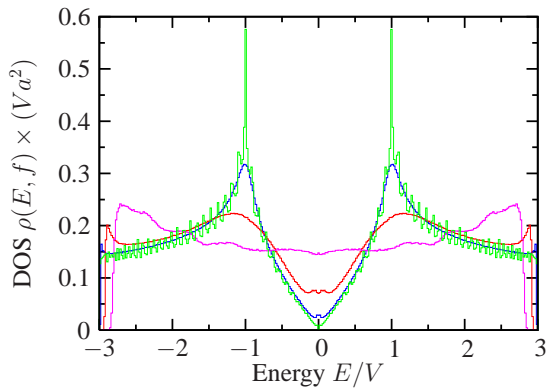}
\caption{(Color online)
The ensemble averaged density of states of a two-dimensional bricklayer
lattice with random flux disorder strength $f=0.05$, 0.2, 0.5, and 1.0. 
With increasing $f$, the van Hove singularities disappear. The
size of the bricklayer system is $L_x\times L_y=64\times 128\,a^2$.  
}
\label{totdos}
\end{figure}

A tight-binding Hamiltonian for non-interacting particles with nearest
neighbor transfer energy $V$ in the presence of perpendicular magnetic fields
is defined by  
\begin{eqnarray}
{\cal H}/V &=& \sum_{x,y}{}^{'} \big(e^{i\theta_{x,y+a;x,y}} c_{x,y}^\dag
c_{x,y+a}\nonumber \\ 
&&
+ e^{-i\theta_{x,y-a;x,y}} c_{x,y}^\dag c_{x,y-a}\big)
\nonumber \\
& &
+ \sum_{x,y} \big(c_{x,y}^\dag c_{x+a,y}+  c_{x,y}^\dag c_{x-a,y}\big),
\label{ham}
\end{eqnarray}
where $c_{x,y}^\dag$ and $c_{x,y}$ are creation and annihilation operators
of a particle at site $(x,y)$, respectively. The prime at the first sum in
(\ref{ham}) indicates that only transfers along the non-zero vertical bonds
are included.  The second sum describes the movement in the horizontal
chains. The phases, which are chosen to be only associated with the vertical 
bonds in the $y$-direction, 
\begin{equation}
   \theta_{x,y;x,y+a}=\theta_{x+2a,y;x+2a,y+a}-\frac{2\pi e}{h}\Phi_{x,y},
\end{equation}
are defined by the total magnetic flux $\Phi_{x,y}=p/q\,(h/e)+\phi_{x,y}$
threading a given plaquette with upper left corner at site $(x,y)$. The 
constant magnetic field is given by the fraction $p/q$ of a flux quantum $h/e$
with mutually prime integers $p$ and $q$ so that the tight-binding band splits
exactly into $2q$ sub-bands, and $\phi_{x,y}$ is the random flux part. 
The latter incorporates the effect of inhomogeneous magnetic fields and
mimics the disorder due to corrugations and ripples\cite{Mea06,GKV08,KG08} 
present in real graphene sheets. In contrast to diagonal disorder, it
preserves the chiral symmetry and ensures a finite conductivity at the Dirac
point.  

\begin{figure}
\includegraphics[width=8.65cm]{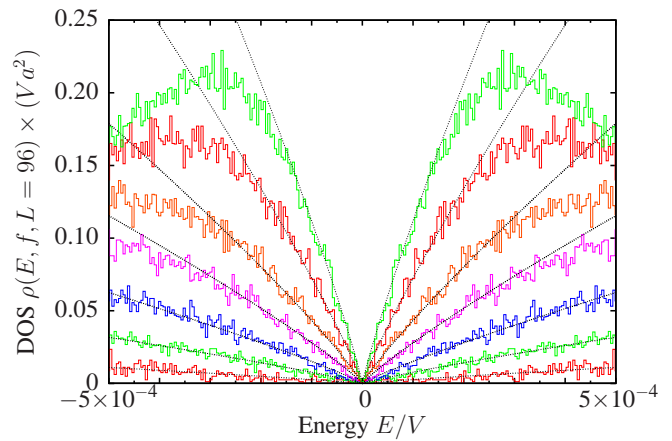}
\caption{(Color online)
The energy dependence of the density of states near the Dirac point with
random flux disorder strength $f/(h/e)=0.2$, 0.3, 0.4, 0.5, 0.6, 0.75, and 1.0. 
The steepest DOS tails belong to $f/(h/e)=1.0$. The system size is $L_x/a=L_y/a=96$.}
\label{closeup}
\end{figure}

The random fluxes are drawn from a box distribution $-f/2\le\phi_{x,y}\le f/2$
with zero mean and disorder strength $0\le f/(h/e)\le 1$. Periodic boundary
conditions are applied in both directions to avoid edge and corner effects and the
system size was chosen to be commensurate with the spatially constant magnetic
field. The eigenvalues $E_i(n)$ of the Hamiltonian (\ref{ham}) were obtained
by direct diagonalization of the $N_r$ disorder realizations and used for the
calculation of the ensemble averaged density of states $\rho(E)$ within an
energy interval $\Delta E$ 
\begin{equation} 
\rho(E)\Delta E=\frac{1}{N_r}\sum_{n=1}^{N_r}\frac{1}{  L_x
  L_y}\int\limits_{E}^{E+\Delta  E} \sum_i \delta (E'-E_i(n))dE'.   
\end{equation}

\section{Density of states}
\subsection{Magnetic field $B=0$}
Starting with the case where the constant part of the magnetic flux density is
zero and only the random disorder part is present,
the ensemble averaged density of states for bricklayer systems of size $L_x
\times L_y = 64 \times 128\,a^2$ is shown in Fig.~\ref{totdos} for different
disorder strengths $f$. With increasing $f$, the sharp van Hove singularities at
$E/V=\pm 1.0$ get rounded and finally disappear. In the same way, the fluctuations,
which can be seen for the smallest disorder $f/(h/e)=0.05$, vanish. The latter 
are due to finite size effects. Based on an energy resolution of 0.02V as used
in Fig.~\ref{totdos}, the other curves do neither depend on the shape of the
system nor on the size which has been checked within the range $32\le L/a\le
192$. The main consequence of the increasing disorder is
seemingly the filling of the valley in the density of states with a strong
increase at the Dirac point $E/V=0$. 
However, a closer inspection of the energy range near the Dirac point reveals a
completely different behavior. As shown in Fig.~\ref{closeup}, independent of
disorder strength, the DOS always goes down to zero at $E/V=0$. 
For small random flux disorder, the DOS vanishes with a slope that finally
becomes $2/(9\pi V^2 a^2)$ in the clean limit. With increasing disorder strength $f$,
this slope becomes steeper and steeper. Since $f/(h/e)=1.0$ is the
strongest random flux disorder possible, there will always be an energy region
around the Dirac point where the density of states goes to zero at $E/V=0$. 

These results have been obtained with different diagonalization methods
including a Lanczos algorithm as well as standard LAPACK routines. 
The number of realizations exceeded $10^4$ for each disorder $f$ and the
DOS bin-width $\Delta E/V$ was $4\times 10^{-6}$.

\begin{figure}
\includegraphics[width=7.0cm]{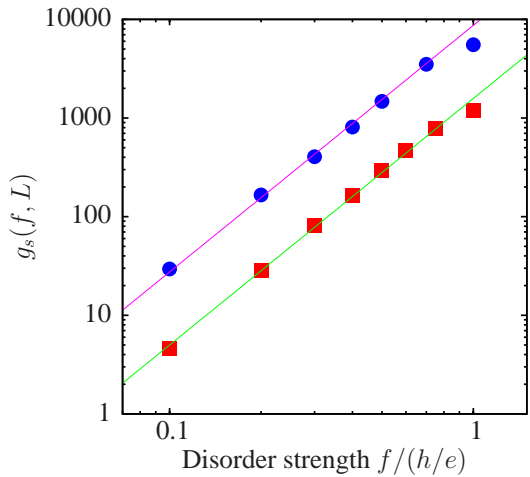}
\caption{(Color online)
The disorder dependence of the fitting parameter $g_s(f,L)$ with $L/a=96$ for
$f/(h/e)=0.1$ and the seven disorder values $f$ shown in Fig.~\ref{closeup}. On
this double-log plot the straight line is given by $g_s(f)=1580\,f^{2.5}$.  
In addition, data for $L/a=192$ are shown with a power-law fit
$g_s(f)=8702\,f^{2.5}$ ($\bullet$). 
}
\label{fdep}
\end{figure}

\begin{figure}
\includegraphics[width=7.5cm]{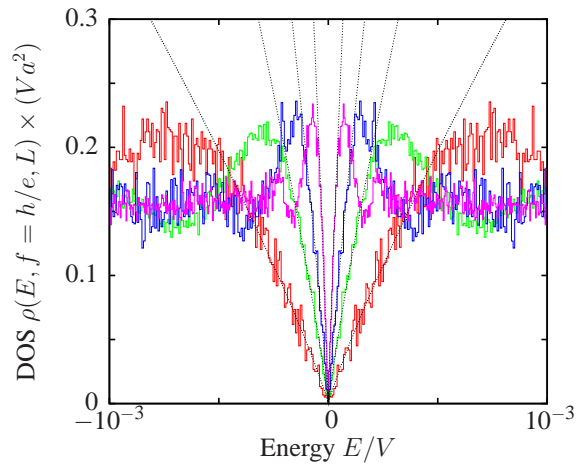}
\caption{(Color online)
The energy dependence of the density of states for disorder $f=h/e$ and system
size $L/a=$64, 96, 128, and 192. With increasing size, the energy range of
the DOS-depression becomes narrower.    
}
\label{Ldepraw}
\end{figure}

A disorder dependent vanishing of the density of states was previously reported for
massless random Dirac fermions on a two-dimensional square lattice\cite{MH97}
and on a honeycomb lattice,\cite{DZT08} where non-diagonal disorder was
introduced by real random hopping terms. These model systems preserve time
reversal symmetry and therefore belong to the (chiral) orthogonal universality
class. Also, the energy range 
where the DOS drops to zero is considerably broader compared with what has
been found in our random flux bricklayer model, which belongs to the chiral
unitary symmetry class. Furthermore, the singular peak observed at $E=0$ in
Ref.~\onlinecite{RH01} for both the disordered random fermions with either
random hopping or random gauge fields is absent in the present bricklayer
situation.     

As seen from the thin black lines used to fit the curves in
Fig.~\ref{closeup}, neither a simple linear energy dependence, as observed in
the clean system, nor a power-law form used in Refs.~\onlinecite{RH01}, \onlinecite{DZT08}
and \onlinecite{MH97} is adequate. An additional logarithmic term similar as 
in the case of the class-D thermal quantum Hall effect\cite{MEMC07,EM08} 
or for Dirac fermions on a honeycomb lattice with weak diagonal and bond
disorder\cite{DZT08} seems to be an appropriate empirical function. Although
bond and random flux disorder are different, both disorder types maintain the
chiral symmetry of the system. Therefore, we try to use the ansatz 
\begin{equation}
\rho(E,f,L)=\frac{|E/V|}{2\pi V a^2}\Big(1+\frac{2}{\pi}g_s(f,L)\ln\frac{1}{|E/V|}\Big),
\label{empir}
\end{equation}
with a disorder and size dependent fitting function $g_s(f,L)$. The latter
grows with both increasing disorder strength and system size $L$. For square 
samples of fixed size $L=(96\,a)^2$ as used in Fig.~\ref{closeup},
a power-law dependence on the disorder strength $g_s(f,L/a=96)\propto f^{2.5}$ 
is found in the range $0.1\le f/(h/e)<1.0$, as is shown in Fig.~\ref{fdep}. 
The same behavior is obtained for larger sizes $L/a=$128 and 192. The
latter data are also included in Fig.~\ref{fdep}.

\begin{figure}
\includegraphics[width=7.0cm]{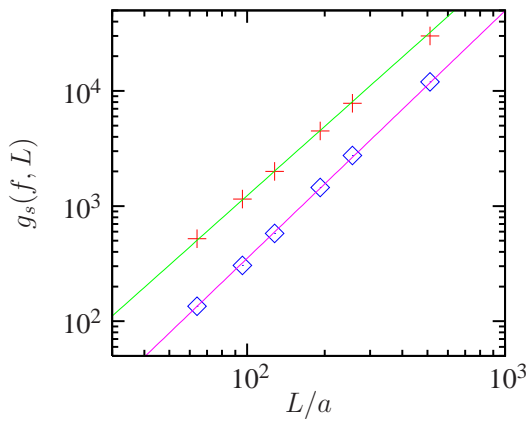}
\caption{(Color online)
The size dependence of the fitting parameter $g_s(f,L)$ for square samples and 
two disorder values $f=1.0\,h/e$ ($+$) and $f=0.5\,h/e$ ($\diamond$). On a
double-log scale, the straight lines are given by
$g_s(f/(h/e)=1.0,L)=0.123\,(L/a)^{2}$ and
$g_s(f/(h/e)=0.5,L)=0.0176\,(L/a)^{2.15}$, respectively.      
}
\label{Ldep}
\end{figure}

The overall shape of the density of states, as plotted in Fig.~\ref{totdos},
seems to be independent of the system size with the exception of noticeable
small finite size fluctuations occurring only for the smallest disorder. In
striking contrast, the depression of the DOS near $E/V=0$ shows a strong length
dependence. In Fig.~\ref{Ldepraw} the averaged density of states for disorder
$f=h/e$ is plotted for sample sizes $L/a=$64, 96, 128, and 192. Applying the
function (\ref{empir}), the size dependence of $g_s$ close to $E/V=0$ can be
obtained. This is shown in Fig.~\ref{Ldep} where the size dependence of the
fitting parameter $g_s(f,L)$ with $64\le L/a\le 512$ is shown on a double-log
plot for two disorder strengths $f=1.0$ and $f=0.5$, respectively. 
In both cases, a power-law relation $g_s(f,L)\propto L^{\kappa}$ is obtained
with an exponent $\kappa=2.0$ for $f=1.0\,h/e$ and $\kappa=2.15$ if
$f=0.5\,h/e$. Because of the uncertainties due to the limited number of
realizations, particularly for larger system sizes, and the restricted range
$64\le L/a\le 512$, it is not possible to rule out that both exponents are the
same in the limit $L\to\infty$.
The energy range where the function (\ref{empir}) can be fitted to the
numerical curves decreases with increasing size $L$ in a similar manner as
with increasing disorder strength $f$, which can be seen in Fig~\ref{closeup}.  

With the disorder and size dependence as identified above from the numerical
data, our ansatz for the density of states (\ref{empir}) used for all
curves shown in Fig.~\ref{closeup}, is the fitting function $g_s(f,L)=g_c
[f/(h/e)]^{5/2} (L/a)^{2}$ with only one adjustable constant $g_c=0.0925$.  

\subsection{Finite magnetic field}
\begin{figure}
\includegraphics[width=7.75cm]{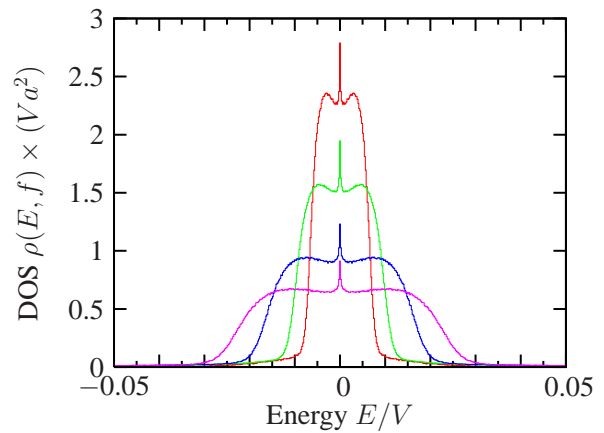}
\caption{(Color online)
The disorder broadening of the lowest (central) Landau band for square samples
of size $L_x=L_y=128\,a$ and constant magnetic flux density $p/q=1/32 (h/e)
a^{-2}$. The disorder strength is $f/(h/e)=0.02, 0.03, 0.05$, and $0.07$,
respectively. 
}
\label{centraLB}
\end{figure}

For a continuum Dirac model in the presence of a finite magnetic field, the
energy spectrum of the charge carriers is arranged into degenerate Landau
levels. The energetically lowest Landau level appears at the charge neutrality
point at $E=0$.\cite{ZA02} Hence, instead of the density of states going to
zero, a finite DOS arises at the Dirac point in the disorder free system for
$B\ne 0$. To account for an external perpendicular B-field in our lattice
model, a spatially constant magnetic  
flux is applied in addition to the random magnetic flux disorder. This type of
disorder causes the central Landau level at the Dirac point to broaden only a
little\cite{SM08a,KHA09} compared with the broadening of the  higher Landau
bands. However, one has to keep in mind that due to the lattice structure, the 
sub-bands exhibit Harper's broadening already in the disorder-free system. This
intrinsic broadening is small and disappears with decreasing magnetic field. 
The disorder broadening of the narrow central Landau band, which is
proportional to $f\sqrt{p/q}$, is seen in Fig.~\ref{centraLB} for $p/q=1/32$
and $L_x=L_y=128\,a$. There is also an additional narrow structure discernible
around $E=0$ and this was previously attributed to originate from the chiral
critical eigenstates.\cite{SM08a}  

\begin{figure}
\includegraphics[width=8.25cm]{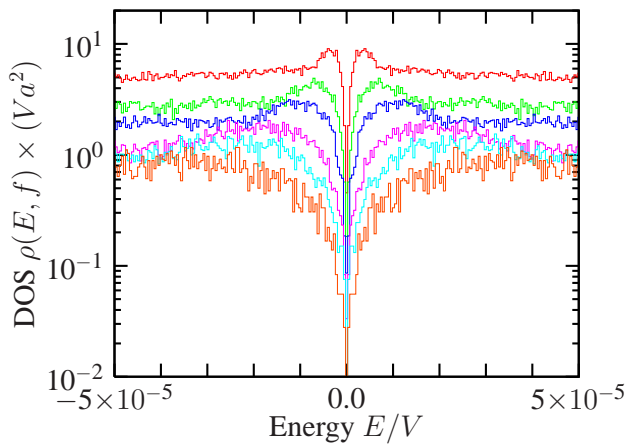}
\caption{(Color online)
The depression of the density of states within the central Landau band near
energy $E/V=0.0$ for square samples of size $L_x=L_y=128\,a$ and constant
magnetic flux density $p/q=1/32 (h/e) a^{-2}$. To enhance the differences, the
DOS is shown on a logarithmic scale. The disorder strength for the various
curves is $f/(h/e)=0.01, 0.02, 0.03, 0.05$, 0.07, and $0.1$, respectively.  
The topmost curve belongs to $f/(h/e)=0.01$.
}
\label{fdep_B}
\end{figure}

In order to scrutinize this special feature, more than $10^4$ disorder
realizations were calculated. Thereby, a bin-width of about $5\times 10^{-7}$
becomes possible leading to an enhanced energy resolution. As a result of
this effort, one can see that the density of states in the center of the
lowest Landau band is by no means constant, but is in fact dominated by a narrow
depression (see Fig.~\ref{fdep_B}), which depends on disorder strength, on the
constant part of the magnetic flux density, and  on the system size. This is
an unexpected outcome and was not identified in previous work. It would be
very interesting to see whether or not a similar depression with a density of
states going to zero at the Dirac point develops also in a continuum Dirac
equation approach. The latter method is generally employed in graphene studies
and believed to be particularly suited near the Dirac point. 

\begin{figure}[b]
\includegraphics[width=7.25cm]{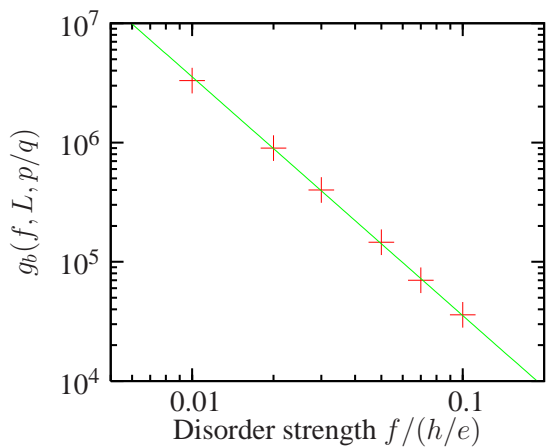}
\caption{(Color online)
The disorder dependence of the fitting parameter $g_b(f,L,p/q)$. The size of
the square samples is $L/a=128$ and the constant magnetic flux density
$B=1/32\,(h/e)a^{-2}$. From the log-log plot, a power-law relation
$g_b(f,L,p/q)=355\,(f/(h/e))^{-2}$ is obtained. 
}
\label{gb_f_dep}
\end{figure}

The density of states within a narrow energy range near $E/V=0$ is shown in
Fig.~\ref{fdep_B} for square samples of size $L/a=128$, magnetic flux density
$B=1/32\,(h/e)a^{-2}$ and several disorder strengths $f$. Contrary to the
zero magnetic field case, the tails flatten with increasing disorder for
finite $B$, and the energy range of the DOS-depression gets broader. Please
note that the DOS is plotted on a logarithmic scale for sake of clarity. 
A similar function as (\ref{empir}) can be used to fit the energy dependence of the
narrow DOS depression, but now with a fitting parameter $g_b(f,L,B)$ which also
depends on the constant magnetic flux density. A power-law relation
$g_b(f,L,B) \propto f^{-2}$ is obtained from the data shown in Fig.~\ref{fdep_B}. 
This behavior is plotted in Fig.~\ref{gb_f_dep} for disorder values in the range
$0.01\le f/(h/e)\le 0.1$. The same dependence has been found also for size
$L/a=192$, magnetic flux $p/q=1/96$ and disorder strengths $f/(h/e)=$0.01,
0.02, and 0.05. Therefore, the energy range of the DOS depression
broadens with increasing disorder strength and the tails become flat, which is
completely opposite to the $B=0$ case.

\begin{figure}
\includegraphics[width=8.25cm]{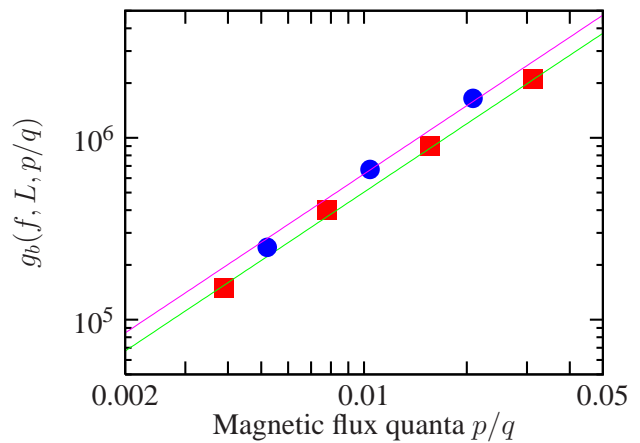}
\caption{(Color online)
The power-law dependence of the fitting parameter $g_b(f,L,p/q)$ versus
magnetic flux quanta $p/q$ for square samples of size $L/a=192$ ($\bullet$) with
$f/(h/e)=0.02$ and for $L/a=256$ with disorder strength $f/(h/e)=0.03$. The
straight lines follow a $\propto (p/q)^{1.25}$ relation. 
}
\label{B-dep}
\end{figure}

The specific magnetic field dependence of $g_b(f,L,B)$ is not so easy to
extract, because both the height and the width of the Landau band vary. In any
case, the DOS tails become steeper with increasing $B$. For square
samples of size $L/a=256$ and disorder strength $f/(h/e)=0.03$, a power-law
relation $g_b(f,L,p/q)\propto (1/q)^{1.25}$ is extracted from a fit to the
relation corresponding to (\ref{empir}) with $q=32, 64, 128$, and 256. 
A similar behavior is observed for $L/a=192$, $f/(h/e)=0.02$, and flux densities
with $q=$48, 96, and 192. Both data fits can be seen in Fig.~\ref{B-dep} on a
double logarithmic scale.  This magnetic field dependence means that
the energy range of the DOS-depression becomes broader and therefore more
important when the magnetic flux density gets smaller, approaching those
applied in experiments. However, this power-law relation will probably only
hold as long as the magnetic length $l_B=\sqrt{\hbar/(eB)}=a\sqrt{q/(2\pi p)}$
remains smaller than the system size $L$.    

The size dependence of $g_b(f=0.01\,h/e,L,p/q=1/32)$ is found to be $\propto
L^{2.5}$ in the range $64\le L/a \le 192$. Although the tails of the
DOS-depression get steeper with increasing system size, $\rho(E/V=0)$ stays
zero at the Dirac point in the range $64\le L\le 512$ investigated. Putting
everything together, in the presence of a perpendicular magnetic field the
dependence of 
the empirical fitting function on disorder strength, system size, and magnetic
flux density can be summarized by $g_b \propto f^{-2} (L/l_B)^{5/2}$.  

\section{Discussion and conclusions}
The density of states of non-interacting electrons moving on a two-dimensional
bricklayer lattice in the presence of chiral symmetry preserving random flux
disorder and a perpendicular magnetic field exhibits a narrow depression near
the Dirac point at $E/V=0$. The corresponding numerical results reveal a
dependence on the disorder strength, the magnetic flux density, and on the
size of the system. The latter is not simply a finite size effect because the
special size dependence develops only near the Dirac point where $\rho(E/V=0)$
stays zero even though the steepness of the tails grows with increasing $L$.  

\begin{figure}
\includegraphics[width=7.5cm]{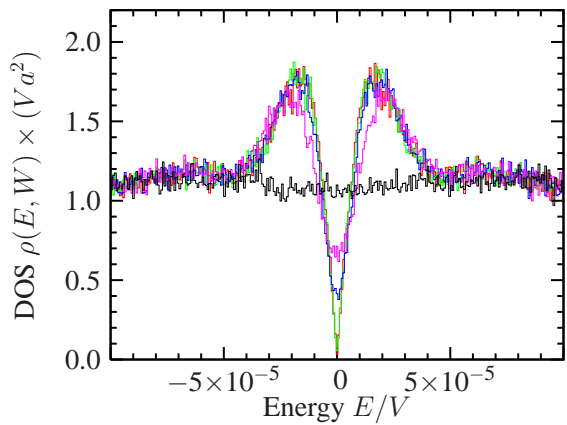}
\caption{(Color online)
The disappearance of the depression in the density of states near the Dirac
point due to an additional diagonal disorder potential of strength
$W/V=10^{-5}, 10^{-4}, 10^{-3}, 2\times 10^{-3}$, and $10^{-2}$, respectively. 
The system size is $L/a=128$, the random flux disorder
$f/(h/e)=0.05$, and the constant magnetic flux density $B=1/32(h/e)a^{-2}$. 
}
\label{diagdis}
\end{figure}

Since the DOS-depression can be removed by an additional diagonal disorder
giving rise to inter-valley scattering, the origin of this feature must
derive from the sub-lattice structure and the associated chiral symmetry of
graphene's honeycomb lattice. 
The modeling of diagonal disorder required an extra term $\sum_{x,y}
\epsilon_{x,y} c_{x,y}^\dag  c_{x,y}$ in the Hamiltonian (\ref{ham}).  
The set of uncorrelated random disorder potentials $\{\epsilon_{x,y}\}$ was
chosen to be box distributed $-W/2 \le \epsilon_{x,y} \le W/2$ with
probability density $1/W$.
The removal of the DOS-depression as a function of additional short-range
disorder potentials can be seen in Fig.~\ref{diagdis} for systems of size
$L/a=128$, random flux disorder $f/(h/e)=0.05$, and magnetic flux density
$B=1/32(h/e)a^{-2}$. With increasing disorder strength $W/V=10^{-5},
10^{-4}, 10^{-3}, 2\times 10^{-3}$, and $10^{-2}$, the
narrow DOS-depression in the lowest (central) Landau band completely disappears. 

The occurrence of the DOS-depression in the absence of short-range diagonal
disorder for $B=0$ could be explained if the elastic scattering length
diverged at the Dirac point. Then, for increasing system size a decreasing
energy range around $E/V=0$ would exist where the elastic scattering length is
larger than $L$. Within this energy interval, the transport would be almost
ballistic and, due to the absence of scattering events, the DOS approaches the
result of the ordered case and drops to zero at $E/V=0$. If this size
dependence of the DOS-depression were accessible in experiments, it would open
the possibility for obtaining information about the elastic scattering length.

Due to the lack of an analytical theory for the density of states of a
disordered bricklayer model near the Dirac point, a relation proposed in the
context of the thermal quantum Hall effect\cite{MEMC07,EM08} and recently for
disordered Dirac fermions on a honeycomb lattice\cite{DZT08} was ventured. 
While the empirical relation (\ref{empir}) used to fit the numerical
results seems to work quite well, one has to keep in mind that only analytical
calculations for a lattice model in the presence of random magnetic flux will
eventually help to understand the complete situation.
The usual way to start from the continuum Dirac equation may turn out to be
not appropriate for a comprehensive description, if the depression in the density
of states found in the present study would not show up in the former description. 

The implication of this observation and its impact on the scaling behavior and
the critical properties at the Dirac point is evident, but still needs to be
investigated. Usually, a noncritical density of states with a smooth energy
dependence is assumed in the scaling analysis. In particular, the strong energy
and size dependence reported above necessitate a reassessment of the
conventional procedure applied in Ref.~\onlinecite{SM08a}.  

Although the disorder and magnetic field dependent depression found in the
density of states at the Dirac point and the occurrence of the conductance
peak splitting reported previously\cite{SM08a} are in agreement with several
aspects observed in experiments mentioned in the introduction, it is clear
that many-body effects and also single particle interactions like Zeeman
splitting or spin-orbit scattering, which were not taken into account in the
present investigations, may turn out to be the dominant effects in
understanding these experiments. Nevertheless, the results of the calculations
presented above may be helpful in finding out which type of disorder
determines the electronic properties of real graphene samples.

\section*{Acknowledgments}
I would like to thank Walter Apel, Ferdinand Evers, and Peter Mark\v{o}s for
helpful discussions.


\end{document}